\documentclass[11pt]{article}
\usepackage{hyperref}
\pdfoutput=1

\begin{document}
\title{Tornadoes in a Microchannel}
\author{Carlos L. Perez and Jonathan D. Posner\\
\\\vspace{6pt} Mechanical and Aerospace Engineering, \\ Arizona State University, Tempe, AZ 85281, USA}
\maketitle

The addition of non-dilute volume fraction of dispersed colloids can significantly alter a fluid's electrical conductivity and permittivity.  Under the application of an external electric field, particles in non-dilute colloidal suspensions can chain and result in unsteady fluid motion, under some conditions.  The application of strong electric fields induces a dipole on the particles.  The induced dipole results in linear particle chaining due to dipolar attraction.  Under higher fields (~400 V cm$^{-1}$) particle chains aggregate assembling into three dimensional structures in the free solution resulting in strong gradients in particle volume fraction.  Gradients in particle volume fraction result in gradients in electrical conductivity and permittivity.  Externally applied electric fields coupled with gradients in electrical conductivity and permittivity result in electric body forces that drive the flow unstable forming vortices. The experiments are conducted in square 200 micron PDMS microfluidic channels.  Colloidal suspensions consisted of 0.01 volume fraction of 2 or 3 micron diameter polystyrene particles in  0.1 mM Phosphate  buffer and  409 mM sucrose to match particle-solution density.  AC electric fields at 20 Hz and strength of 430 to 600 V cm$^{-1}$ were  used. We present a fluid dynamics \href{http://hdl.handle.net/1813/11494}{video} that shows the evolution of the particle aggregation and formation of vortical flow.  Upon application of the field particles aggregate forming particle chains and three dimensional structures.  These particles form rotating bands where the axis of rotation varies with time and can collide with other rotating bands forming increasingly larger bands. Some groups become vortices with a stable axis of rotation. Other phenomena showed include counter rotating vortices, colliding vortices, and non-rotating particle bands with internal waves.

\end{document}